\documentclass[aps]{revtex4}
\usepackage{amsfonts}
\usepackage{amssymb}
\usepackage{graphicx}%

\newcommand{\beq}{\begin{equation}}
\newcommand{\eeq}{\end{equation}}
\newcommand{\beqn}{\begin{eqnarray}}
\newcommand{\eeqn}{\end{eqnarray}}
\newcommand{\bearr}{\begin{array}}
\newcommand{\enarr}{\end{array}}

\begin{document}


\title{A new approach to partial synchronization in globally coupled
rotators}

\author{P. K. Mohanty}  
\email[ Current Address : TCMP Division,
 Saha Institute of Nuclear Physics,
 1/AF Bidhan Nagar, Kolkata-700 064, India\\ E-mail : pk.mohanty@saha.ac.in]{}
\author{Antonio Politi}
\email[
Permanent address:
CNR -- Istituto dei Sistemi Complessi, via Madonna del Piano, 10 I-50019
Sesto Fiorentino Italy\\E-mail : antonio.politi@isc.cnr.it] {}

\affiliation{Max-Planck Institute f$\ddot{u}$r Physik Komplexer Systeme,
N$\ddot{o}$thnitzer Str. 38, D-01187 Dresden, Germany}
 
\date{\today}
\begin{abstract}
We develop a formalism to analyze the behaviour of pulse--coupled identical
phase oscillators with a specific attention devoted to the onset of partial
synchronization. The method, which allows describing the dynamics both at the
microscopic and macroscopic level, is introduced in a general context, but then
the application to the dynamics of leaky integrate-and-fire (LIF) neurons is
analysed. As a result, we derive a set of delayed equations describing 
exactly the LIF behaviour in the thermodynamic limit. We also investigate the 
weak coupling regime by means of a perturbative analysis, which reveals that 
the evolution rule reduces to a set of ordinary differential equations. 
Robustness and generality of the partial synchronization regime is finally 
tested both by adding noise and considering different force fields.
\end{abstract}
\maketitle

\section{Introduction}
Understanding the behavior of networks of dynamical units is a very important
subject of research in various contexts such as information processing in the
brain, or metabolic systems. Given the richness of the observed phenomenology
and the relevance of many different ingredients (e.g., the topology of the
connections, the presence of disorder and of delayed interactions), it is very
instructive to consider even simple setups in the perspective of
identifying the hopefully few mechanisms responsible for the main general
phenomena. Here, we focus our attention on a type of collective behavior
arising in globally coupled identical systems.
The first evidence of collective behavior dates back to the 70's, when it was
proven that a nonzero mean field may spontaneously arise in an ensemble of
particles stochastically moving in a bistable potential \cite{KS75}. Later, it
was numerically shown that macroscopic periodic dynamics appears in coupled
stochastic \cite{BCM87} as well as chaotic (R\"ossler) oscillators \cite{PRK96},
while the first experimental evidence was found in Josephson--junction
arrays \cite{WCS96}. Collective behaviour may also arise in the presence of
quenched disorder as revealed by the Kuramoto model~\cite{K84},
introduced to explain the onset of synchronization in an ensemble of different
units~\cite{Kbook}. Over the years it has been discovered that collective 
dynamics can be also chaotic, as found in globally coupled maps \cite{K90} and 
oscillators \cite{GHSS92}.

Here we investigate ``partial synchronization'' (pS), a little known
phenomenon discovered in pulse--coupled, leaky integrate-and-fire (LIF)
neurons \cite{V96}. In this regime, the mean field exhibits a periodic dynamics
which, at variance with other contexts, arises in the absence of any
synchronization between the single units which behave quasi-periodically. 
pS arises from the destabilization of a regime characterized by a constant mean
field and a periodic behaviour of the single elements, their phases being
equispaced. This ``homogeneous regime" has been found in many different
contexts such as Josephson devices \cite{HB87}, multi-mode lasers \cite{WBJR90}
and electronic circuits~\cite{AKS90}. By following Ref.~\cite{NW92}, it is now
called ``splay state'' and its microscopic stability properties have been
studied  quite in detail in~\cite{SM93}. Since splay states are rather general,
it is reasonable to expect that the transition to pS can be observed in a wide 
class of systems too. In fact, the preliminary evidence of pS found in
Hindmarsh-Rose  neurons and van der Pol oscillators \cite{PR} confirms such
expectations and poses the question of understanding the general conditions
for the onset of pS.

In order to unravel this point, we develop a new formalism that allows treating
pulse coupled oscillators and establishes a possible basis for an extension to
different coupling schemes. The approach is introduced in a general framework 
to show the potentiality of the method, but the resulting delayed equations 
are analyzed only in the specific case of LIF
neurons, since in this case a simple and explicit expression is available for
one of the relevant variables: the ``finite-time" Lyapunov exponent $\Lambda$.
Furthermore, in the attempt of identifying the minimal model able to describe
the onset of pS, we have explored both analytically and numerically the weak
coupling limit, showing that a perturbative expansion allows reducing the
delayed to ordinary equations.
The persistence of the transition to pS for an arbitrarily small coupling
strength paves the way towards more general setups. A further
promising indication is given by the numerical observation that pS arises also
for continuous force fields, which again suggests that the phenomenon is more
general than initially conjectured \cite{V96}. Finally, we have added noise to the
single neuron dynamics, finding that the macroscopic oscillations persist,
though with a smaller amplitude.

\section{Modelling identical rotators}
The investigation of globally coupled systems requires setting up a suitable
(nonlinear) self-consistent approach. A much used approach to describe the
collective behaviour of neural networks is dynamical mean field
\cite{T93,GV93,G95,B99}. Here, in the absence of noise and of microscopic chaos,
there is no justification a priori for coarse-graining and indeed the
transition to pS can be observed already in ensembles of finitely many
oscillators \cite{V96} and can thus also be viewed as a bifurcation. It is
therefore desirable to develop a formalism to capture both micro- and
macro-scopic features. We start from the crossing-times as they are the most
suitable variables to characterize splay states. Our approach can, in fact,
be viewed as an extension of the method followed in Ref.\cite{AV93}, where
pseudo-crossing times have been introduced with reference to the steady
dynamics. On the other hand, it is fair to recognize that the final model
equations are equivalent to those obtained by implementing a dynamical mean
field approach, when the latter is applicable. This implies that in the present
context one can smoothly go from the microscopic to the macroscopic level.

We consider an ensemble of $N$ phase oscillators (i.e. rotators), each one
characterized by the phase
$\phi_i$, $i=0,\ldots N-1$  which evolves according to the equation
\begin{equation}
\dot \phi_i = F(\phi_i,E(t))
\label{eq:velocity}
\end{equation}
where $E$ is a suitable mean field whose dynamics depends homogeneously on all
the $\phi_i$ in a way that is not crucial for the present discussion.
The force field $F(\phi,E)$ is assumed to be positive and periodic in $\phi$
with period 1, but it can be discontinuous as, e.g., a sawtooth function
(this is indeed the case considered in the next section).
We start defining a suitable Poincar\'e section, by introducing a
threshold $\overline \phi$, which, without loss of generality, is set in
$\phi=1$. Let then $t_m$ and $i(m)$ denote, respectively, the $m$th crossing 
time ($m=1, \ldots, \infty$) of the threshold and the corresponding rotator 
crossing the threshold. Once the labels $i$ are ordered in such
a way that $\phi_{i+1}<\phi_i$, it follows that $i(m+N) = i(m){\rm mod}~N$ 
\footnote{According to Eq.~(\ref{eq:velocity}), phases cannot cross each 
other, since two oscillators with the same phase must have also the same 
velocity. Therefore, the threshold is always crossed in the same order.}
$i.e.$, the same oscillator crosses the threshold at time $t_m$ and $t_{m+N}$.

The core of the approach consists in deriving an equation linking
$\delta t_m \equiv t_{m+1} - t_{m}$ with $\delta t_{m-N}$
in the limit of large $N$. Simple arguments show that, up to higher-order
corrections in $1/N$,
\begin{equation}\label{core}
\hskip -2.cm \delta t_m = \frac{\delta \phi_{i(m)}(t_{m})}{F(1,E(t_m))} \frac {\delta \phi_{i(m)}(t_{m-N})}{F(1,E(t_m))} {\rm e}^{\Lambda(\{E\})T_m}
= \delta t_{m-N} \frac {F(0,E(t_{m-N}))}{F(1,E(t_{m}))} {\rm e}^{\Lambda(\{E\})T_m}
\end{equation} 
where $\delta \phi_j \equiv \phi_j -\phi_{j+1}$
(here $j$ is meant modulo $N$), while $T_m = t_m - t_{m-N}$ is
the interval between two consecutive threshold crossings of the same oscillator;
finally, $\Lambda(\{E\})$ is the finite-time Lyapunov exponent for a trajectory
starting at time $t_{m-N}$ in $\phi=0$ and ending at time $t_m$ in $\phi=1$, and
thus, in general, depends on the dynamics of the field $E$. The first and last
equalities in the above equation state simply that the temporal separation is
equal to the phase separation divided by the instantaneous velocity. The middle
equality follows from the observation that, being $\delta \phi_{i(m)}$ a small
quantity, it evolves according to the linearized dynamics. In view of the
possible existence of a discontinuity in the force field, we carefully 
distinguish between the $\phi$ value just before ($\phi=1$) and
after $(\phi=0$) the threshold crossing.

If the crossing-time distribution is smooth (this assumption can be verified a
posteriori by integrating the resulting model), then $\delta t_m$ has to be of
the order of $1/N$ and one can accordingly introduce the ``instantaneous" flux
\begin{equation}
\pi(t) \equiv \frac{1/N}{\delta t_m} ,
\end{equation}
where we have dropped the subscript $m$ in the l.h.s. as in the limit
$N\to \infty$, the time variable becomes again continuous. 
Accordingly, Eq.~(\ref{core}) simplifies to
\begin{equation}
\pi(t) = \pi(t-T) \frac {F(1,E(t))}{F(0,E(t-T))} {\rm e}^{-\Lambda(\{E\})T} \, ,
\label{eq:model1}
\end{equation}
where the time interval $T$ is self-consistently determined from the integral 
expression
\begin{equation}
\int_{t-T(t)}^t dt^\prime \pi(t^\prime) = 1 \quad ,
\label{eq:int_pi}
\end{equation}
expressing the condition that from time $t-T(t)$ to time $t$, all oscillators
cross the threshold. An equivalent and more appealing equation is obtained by 
evaluating the time derivative of Eq.~(\ref{eq:int_pi}), namely
\begin{equation}
\dot T = 1 - \frac{\pi(t)}{\pi(t-T)} .
\label{eq:model2}
\end{equation}
In order to complete the derivation of a closed set of equations, it is
necessary to include the mean field dynamics. In the context of pulse coupled
systems, this can be easily done, since $E$ is, by definition, the linear
superposition of the pulses emitted by the single oscillators whenever they 
reach a threshold that, without loss of generality, can be assumed to 
coincide with
$\overline \phi$. Accordingly, the mean field satisfies a linear
differential equation of the type
\begin{equation}
{\mathcal L}(E,\dot E, \ddot E, \ldots) = \pi(t) ,
\label{eq:model3}
\end{equation}
where the firing rate $\pi(t)$ is nothing but the previously introduced
instantaneous flux, while the structure of the operator $\mathcal L$ can be
determined by imposing that the corresponding Green function coincides with
the assumed shape of the emitted pulses.
This completes the derivation of the final equations that are given
by Eqs.~(\ref{eq:model1},\ref{eq:model2},\ref{eq:model3}). The model combines
properties of discrete (Eq.~(\ref{eq:model1})) and continuous
(Eqs.~(\ref{eq:model2},\ref{eq:model3})) time systems, besides involving a
time-dependent self-adjusting delay. Except for the first property, its
structure resembles that of threshold delay equations \cite{LER01} introduced,
e.g., in epidemiology and immunology (see \cite{SK92} for a similar model).

\section{An example: LIF neurons}

In this section we specifically refer to the ensemble of LIF neurons studied
already in Ref.~\cite{V96}. In this case, $\phi$ corresponds to the (adimensional)
membrane potential which evolves according to the force field
\begin{equation}
F(\phi,E) = a + \lambda(1-\phi) + g E(t)  \, ,
\label{eq:model3b}
\end{equation}
where $g$ is the coupling constant, while $a$ and $\lambda$ determine the
dependence of the force field on the potential $\phi$. If $\phi$ reaches the
threshold $\overline \phi=1$ at time $t_0$, a pulse $p(t-t_0)$ is emitted (and
received by all neurons), while the potential is reset to 0. The resetting
makes it legitimate to interpret $\phi$ as a phase variable, since one can
formally extend $\phi$ to the whole real axis by identifying $\phi$ with
$\phi+1$ and thereby see the evolution as a continuous process.
Paying attention to the discontinuities present in the force field,
Eq.~(\ref{eq:model1}) writes as
\begin{equation}
\pi(t) = \pi(t-T) \frac {a + g E(t)}{a + \lambda +g E(t-T)} {\rm e}^{\lambda T}
\quad ,
\label{eq:model1a}
\end{equation}
where we have also taken into account that, because of the linearity of 
the force field, the {\it Lyapunov exponent} $\Lambda$ appearing in
Eq.~(\ref{eq:model1}) is equal to $-\lambda$ independently of the $E$ dynamics.

As for the mean field, if we assume that the single pulse shape is
$p(t-t_0)= \alpha^2(t-t_0){\rm e}^{-\alpha(t-t_0)}$, one finds that $E$
satisfies the equation \cite{V96}
\begin{equation}
  \ddot E + 2\alpha \dot E + \alpha^2 E = \alpha^2 \pi  \quad .
\label{eq:model3a}
\end{equation}

As a result, the model reduces to the set of equations
(\ref{eq:model2},\ref{eq:model1a},\ref{eq:model3a}). For a small enough $\alpha$
value, the dynamics converges to a fixed point.
Since all neurons sequentially cross the firing threshold, the translational 
invariance of a fixed point automatically implies a uniform distribution of the
phase differences, that is the distinguishing feature of splay states.
A tedious but strightforward linear stability analysis proves that the splay
state destabilizes at the critical value already identified in \cite{V96}, where
a Hopf bifurcation signals theonset of a collective periodic behaviour, i.e. of
pS.

\begin{figure}
\begin{center}

\includegraphics*[width=6.0cm,height=5 cm]{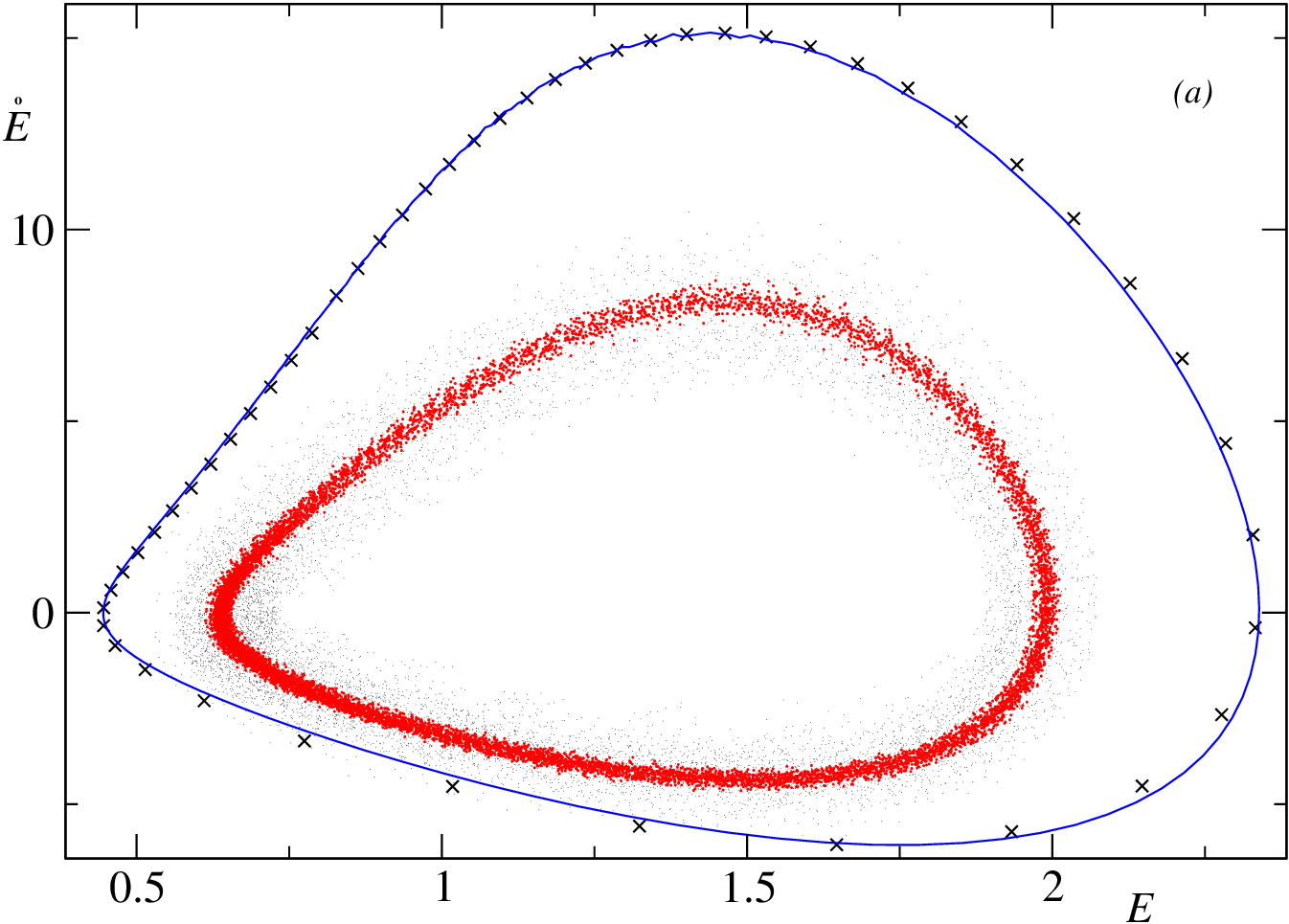}
\includegraphics*[width=7.5 cm,height=5 cm]{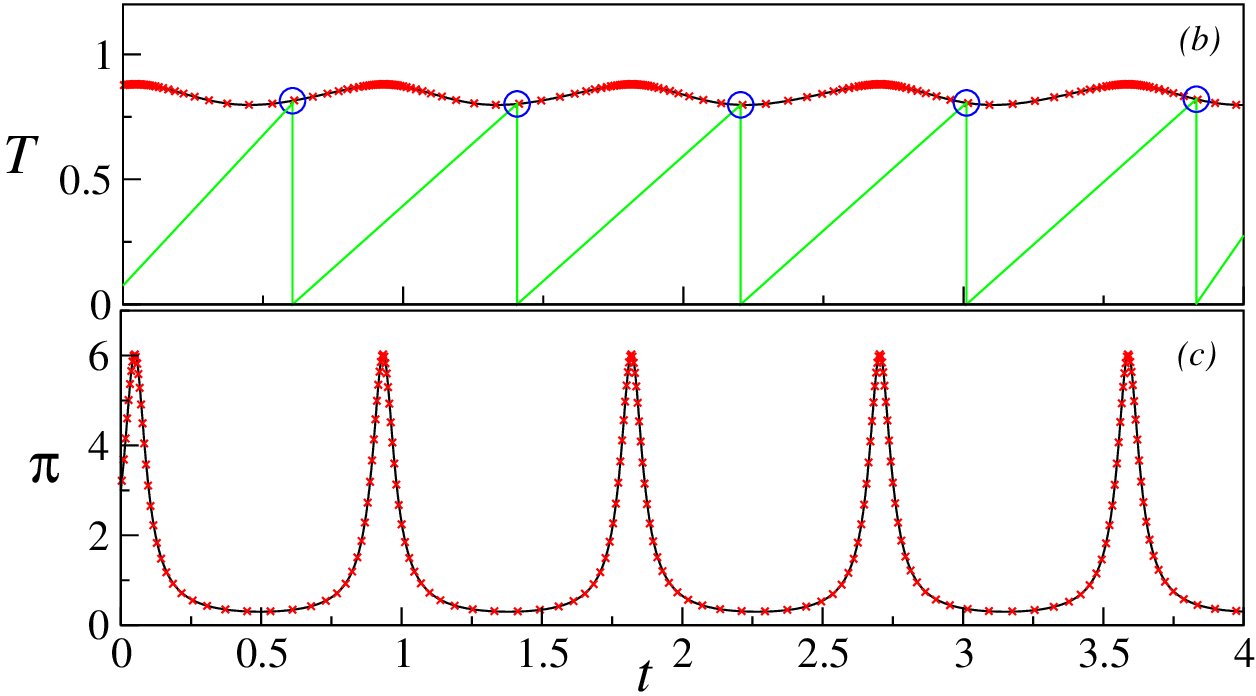}

\caption{Comparison between the initial model for $N=100$ oscillators (crosses)
and the set of equations (\ref{eq:model2},\ref{eq:model1a},\ref{eq:model3a})
(solid line) for $a=0.3$, $g=0.4$, $\lambda=1$ and $\alpha=9$. The
representation in the phase space $(\dot E,E)$ is plotted in panel (a), while
the time evolution of $T$ and $\pi$ is drawn in (b) and (c). The dots in panel
(a) correspond to simulations performed in the presence of additive uniformly
distributed noise of width  $\sigma = 0.025$ (the wider band corresponds to
$N=200$, while the other to $N=4000$). Finally, a geometric procedure for
reconstructing the single neuron firing times is plotted in (b).}
\label{fig:ET}
\end{center}

\end{figure}

In Fig.~\ref{fig:ET} we present the outcome of numerical simulations carried on
in the pS regime. First of all, notice that the results are basically
indistinguishable from those of the original set of equations, as they should.
It is then instructive to look at $T(t)$, since it provides a bridge with the
microscopic description. In fact, if we denote with $t_n[i]$ the time when the
$i$th neuron emits a spike (i.e., it crosses the threshold $\overline \phi$),
$T(t_n[i])$ represents the last inter-spike interval (ISI) of the very same
neuron and all past spike times are obtained by simply iterating the recursive
formula 
\begin{equation}
t_{n-1}[i] = t_n[i]- T(t_n[i]).
\label{eq:rotor}
\end{equation}
Geometrically, the implemetation of this formula corresponds to the 
following procedure (see Fig.~\ref{fig:ET}b): given the point $(t_n,T(t_n))$, 
move down along a straight line of slope $1$ until the $t$-axis is reached at 
the time $t_{n-1}$ and then vertically up until the $T$-curve is hit. 
Considering that $T(t)$ is a positive definite periodic function,
Eq.~(\ref{eq:rotor}) describes nothing but the evolution equation of a
periodically forced oscillator, so that one expects the onset of locking
phenomena. However, it appears that the self-generated $T$ dynamics does not
exhibit any such locking, even when the ratio of the two frequencies is
rational. This remarkable feature remains unexplained.

Another peculiarity of pS is that the single--neuron ISI differs from (it is
always smaller than) the period of the macroscopic dynamics. Interestingly
enough, this is the distinguishing feature of slowly oscillating periodic
solutions already found in equations with state-dependent delay \cite{LER01}.
Qualitatively, the phenomenon can be clarified by noticing that neurons tend to
bunch together, but, at the same time, those neurons lying in the front of the
cluster escape away, while those reaching the back get stuck. This justifies
the name ``partial synchronization'' attributed to this phenomenon.

\section{Perturbative analysis}
\begin{figure}
\centerline{\includegraphics*[width=7.5 cm]{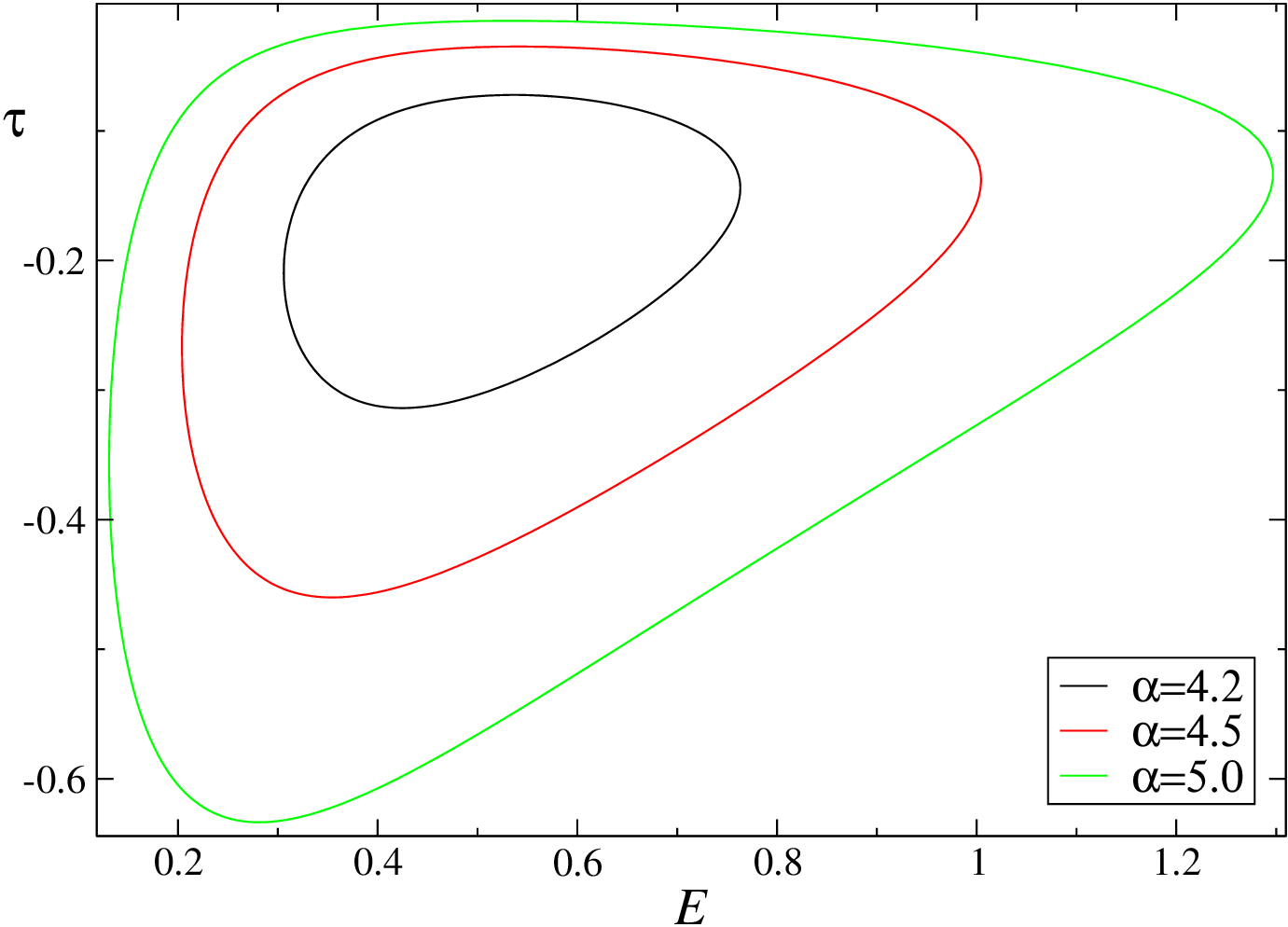}}
\caption{The amplitude of $E$ and $\tau$ decreases as one approaches
$\alpha_c=4.055(8)$ (the critical point for $\lambda=0.001$ 
and $g=0.4$). The transition is thus a supercritical Hopf bifurcation.}
\label{fig:deltaE}
\end{figure}

Upon exploring the parameter plane $(\alpha,\lambda)$, we have discovered that
pS can be observed for arbitrarily small $\lambda$-values. As a result, one can
hope to shed further light on this phenomenon, by developing a perturbative
analysis. This is not easy, because the limit case $\lambda=0$ is degenerate, as
Eq.~(\ref{eq:model1a}) is satisfied by any periodic function $\pi(t)$. In fact,
in this limit, there are no effective interactions among the oscillators and
their distribution does not change in time, whatever the initial choice is. The
only constraints are given by Eq.~(\ref{eq:model3a}), which allows determining
$E(t)$ and Eq.~(\ref{eq:int_pi}), which fixes the period $P=(1-g)/a$. Thus,
exactly for $\lambda = 0$, pS cannot arise, since the periodicity of
the single oscillators necessarily coincides with the collective periodicity.
However, as soon as $\lambda>0$, $T$ can differ from the period $P$,
$T(t)= P + \lambda \tau(t)$. Accordingly, for a generic function $u(t)$ periodic
of period $P$, one can write
\begin{equation}
u(t-T) = u(t-\lambda \tau) = u(t)- \lambda \tau \dot u(t) + O(\lambda^2).
\label{eq:uT}
\end{equation}
By introducing this expansion into the model equations
(\ref{eq:model1a},\ref{eq:model3a}), one obtains, up to first order in
$\lambda$,
\begin{eqnarray}
&&\dot \tau = -P + \frac{1-g\tau \dot E}{ a+gE} \cr
&& \ddot E + 2\alpha \dot E +  \alpha^2 E =  \alpha^2 \frac{C}{\tau} .
\label{eq:lambda0}
\end{eqnarray} 
The same procedure, when applied to Eq.~(\ref{eq:model2}), implies that
$\pi \tau=C$, where $C$ is a constant of motion, so that the model reduces to a
simple set of ordinary differential equations in a three dimensional space
($E$, $\dot E$ and $\tau$ being the relevant variables). The a priori unknown
parameter $C$ must be determined self-consistently by imposing that the integral
of $\pi = C/\tau$ over the period $P$ is equal to $1$. By numerically integrating
the above set of equations, we find a self-consistent solution above
a suitable critical $\alpha$-value. The corresponding collective
behaviour is plotted in Fig.~\ref{fig:deltaE} for different $\alpha$-values. 

\section{Conclusions and perspectives}
In order to test the robustness of the pS regime, we have performed some
further studies both to verify whether the discontinuity in the force
field is a necessary condition and to explore the stability against the
addition of external noise. The simplest way to introduce a continuous force
field is by adding a branch $F'(\phi) = a + \lambda' \phi + E$ (with
$\phi<\phi_0= \lambda/(\lambda'-\lambda)$) to Eq.~(\ref{eq:model3b}) which is 
now restricted to $\phi>\phi_0$. It is in principle possible to follow the same
approach that has led to Eq.~(\ref{eq:model1a}), but here we limit ourselves to
mentioning the result of our numerical simulations, namely that for
$\lambda' >\approx 14 \lambda$, the continuous model keeps exhibiting a
transition to pS. Finally, we have numerically analysed the original model,
under the further action of a zero-average uniformly distributed noise of
width $\sigma = 0.025$ for $N=200$ and $4000$. The results, plotted in
Fig.~\ref{fig:ET}, show that the collective behaviour, though depressed by the
noise action, persists. In fact, the transversal fuzziness appears to decrease
with the number $N$ of oscillators as $1/\sqrt{N}$.

In the second section we have obtained a closed set of equations to treat
identical pulse-coupled rotators. The main obstacle towards specific
applications is the need to express the finite-time Lyapunov exponent
in Eq.~(\ref{eq:model1}) in terms of the other relevant variables. 
This seems to be
only a technical problem, that we have indeed recently solved for generic
sinusoidal fields \cite{prog}. On the other hand, it is less clear how to
incorporate the effect of noise and disorder. In this respect, a close
comparison with the dynamical mean field approach should be very helpful,
since noise can be easily handled by the latter one, while our
method is more open to the treatment of generic force fields.

We thank an unknown referee for having drawn to our attention a similar
bifurcation found in systems with variable time delay.

\end{document}